# Title: Ramsey interference with single photons

**Authors:** Stéphane Clemmen[1*], Alessandro Farsi[1], Sven Ramelow[1,2] and Alexander L. Gaeta[1,3]

**Affiliations:**

[1]School of Applied and Engineering Physics, Cornell University, Ithaca, New York 14853, USA

[2]Faculty of Physics, University of Vienna, 1090 Vienna, Austria

[3]Kavli Institute at Cornell for Nanoscale Science, Cornell University, Ithaca, New York 14853, USA

*Correspondence to:  sclemmen@ulb.ac.be

**Abstract**: Interferometry using discrete energy levels in nuclear, atomic or molecular systems is the foundation for a wide range of physical phenomena and enables powerful techniques such as nuclear magnetic resonance, electron spin resonance, Ramsey-based spectroscopy and laser/maser technology. It also plays a unique role in quantum information processing as qubits are realized as energy superposition states of single quantum systems. Here, we demonstrate quantum interference of different energy states of single quanta of light in full analogy to energy levels of atoms or nuclear spins and implement a Ramsey interferometer with single photons. We experimentally generate energy superposition states of a single photon and manipulate them with unitary transformations to realize arbitrary projective measurements, which allows for the realization a high-visibility single-photon Ramsey interferometer.  Our approach opens the path for frequency-encoded photonic qubits in quantum information processing and quantum communication.

**Main Text:** The two-state model represents the most fundamental quantum system and can be applied to a wide variety of physical systems. Ramsey interferometry, magnetic resonance imaging, and electron-spin resonance spectroscopy are governed by similar 2-level system dynamics, which involves molecular-atomic levels, nuclear spin, and electronic spin, respectively. The coupling between energy levels is achieved using electromagnetic fields, which can be tailored at will and allows for many advanced techniques such as adiabatic elimination and stimulated Raman adiabatic passage in higher dimensional atomic system, or spin locking in NMR.  Quantum interference involving systems in superposition of different energies is at the heart of fundamental and applied physics.  For example, quantum coherence has been highly useful in increasing the accuracy of time measurement from the first idea of using NMR suggested by Rabi in 1945 (*1*) to the first atomic clock relying on the Ramsey interferometry (*2,3*), which has been recently extended by using trapped single ions (*4*). In addition, Ramsey interferometry on single Rydberg atoms has allowed the nondestructive measurement of the number of photon in a cavity (*5*) and single spin manipulation using the same techniques constitutes one of the most promising routes towards quantum processing (*6-8*). Matter-wave interferometers using collective energy levels of atoms in a BEC have also been demonstrated (*9*) and used to measure gravity down to record breaking precision (*10*). Nevertheless, a fundamental quantum system that has not been extensively studied in the context of discrete 2-level energy systems (i.e. frequency) is the single photon. Translating those studies to photonics system can be implemented by controlling light with light using nonlinear optics. For classical light the analogy between atomic/molecular optics and nonlinear optics is well known (*11*) and there are various cases where the complex dynamics of light propagation in a nonlinear medium can be simplified to the coherent evolution of a two-level system. For a quanta of light a bichromatic qubit is a photon whose frequency can be one of two possible colors. A key requirement is to manipulate the frequency states of single photons while preserving their coherence such that transitions between the two frequencies mimic Rabi oscillations.

In this report, we demonstrate manipulation of single photons and their corresponding position on the energy Bloch sphere and show the resulting quantum interference associated with two-level quantum systems. On the Bloch sphere, polar rotations are achieved using a phase sensitive wave-mixing process known as Bragg scattering four-wave mixing (BS-FWM) which translates the frequency of the initial state to a new frequency without adding noise. Azimuthal rotations are implemented by imparting tunable physical delays on the single photon. This approach to encoding quantum information onto the energy degree of freedom of single photons represents an important advance on frequency-encoded qubits since in our case such a qubit is clearly implemented and defined. Such a representation was not straightforward to realize in earlier works (*12*). It has been shown that photons from different frequencies interfere at the single quantum level (*13*) and that such two-parties system can violate Bells inequalities (*14*). However, such studies do not allow quantum logic operations in a 2-dimensional Hilbert space, and a more complex definition of a frequency-encoded qubit is required (*15*) to realize this ideal case. The core of our work is the efficient and low-noise frequency conversion of single photons that was proposed 2 decades ago (*16*) and for which various successful studies using $\chi^{(2)}$ and $\chi^{(3)}$ nonlinear interactions have been demonstrated (*17-28*). It is also interesting to note that similar ideas are being developed for the purpose of sorting optical temporal modes (*29*).

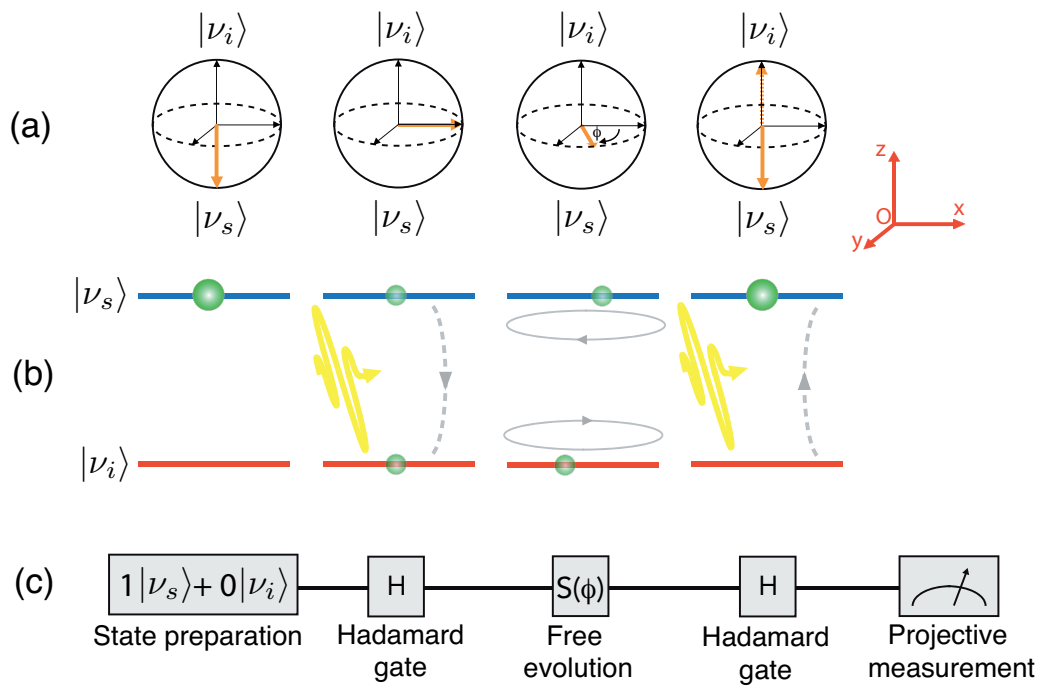

Figure 1. Principle of a Ramsey interference based on discrete energy levels of photons depicted as (a) rotations on the Bloch sphere, (b) evolution of a population on a two level system, or (c) in term of quantum logic.

## Principle and implementation

Figure 1 depicts the minimum set of functions required to perform quantum logic based on two energy levels of a single photon. This can be represented as a Ramsey interferometer and described either in terms of quantum logic operations or as population evolution based on rotations on a Block sphere where the poles represent states of the two different energy levels. In a Ramsey interferometer, a two-level system undergoes subsequent identical interactions with resonant pulses of area $\pi/2$ that are separated by a non-interacting interval of free evolution. For the photonic realization we first define two discrete frequencies and prepare the photon in one, which we refer to as $|v_s\rangle$ that corresponds to being on the south pole of the Bloch sphere. A $\pi/2$ pulse transforms this pure state into a superposition of the form $2^{-1/2}(|v_s\rangle + |v_i\rangle)$, which corresponds to a $\pi/2$ polar rotation in the $O_{x,z}$ plane along the meridian of the Bloch sphere. The bichromatic qubit is subsequently left free to evolve for a time $T$ resulting in an azimuthal rotation, which corresponds to the system acquiring a relative phase $\varphi = 2\pi\delta vT$ due to the precession between the to two levels of different frequency $\delta v$. Finally, a second $\pi/2$ pulse is applied that transforms the superposition state into the final state which depends critically on the imparted phase $\varphi$. The projective measurement consists of detecting whether the photon has a frequency $v_s$ or $v_i$ and therefore reveals information on the phase $\varphi$.

In order to prepare any state of a bichromatic qubit, that is, implement the scheme depicted in fig. 1, our toolbox requires 4 elements: (A) a bichromatic qubit defined by 2-dimensional Hilbert space $\{|v_s\rangle,|v_i\rangle\}$, (B) a photon frequency converter capable of transferring the eigenvector back and forth ($\pi/2$ pulse), (C) control on the relative phase $\varphi$, and (D) a measurement of the energy of the final state.

To define a bichromatic qubit (A) and initialize it on a pole of the Bloch sphere, we isolate a single photon at a given frequency (A) $v_s$ by using a frequency heralded photon source. Photon pairs are generated via spontaneous down conversion over frequencies $v_s$ and $v_{heralding}$ so that energy is conserved $v_s + v_{heralding} = v_{pump}$ where $v_{pump}$ is the fixed frequency of a pump beam. By spectrally filtering the broad flux of photons at the frequency $v_{heralding}$ prior to its detection in a single-photon detector, the partner photon at frequency $v_s$ is characterized in time and frequency. Indeed, as heralding and heralded photons are created as a single event, the timing of heralded photon is known down to the timing resolution of the heralding detector while its spectrum is bounded by the bandwidth of the bandpass filter placed in front of the heralding detector and the energy conservation.

To manipulate the bichromatic qubit we perform quantum frequency conversion (QFC) on the generated single photons. QFC has been demonstrated using second order $\chi^{(2)}$ nonlinearities either via sum-frequency generation *(16,17,22-27)* or electro-optic modulation *(13–15,30)*. In contrast, Bragg scattering four-wave mixing is a third order $\chi^{(3)}$ nonlinear process that has also been shown to be an effective *(18–21)* way of achieving QFC which can be applied in many types of waveguide platforms. A major advantage of BS-FWM over sum-frequency generation and electro-optic modulation is its reasonably little constraint concerning the frequencies involved in the conversion. Indeed, for BS-FWM the principal constraint lies in the phase matching condition while sum frequency generation has the additional requirement that interacting fields have to be at least one octave apart and electro-optic modulation is bounded to the other extreme to frequency shift in the GHz range because of the electric modulation involved.

In our experiment we employ a BS-FWM configuration that is depicted in Fig. 2 in which two strong fields $E_1$ and $E_2$ (referred to as pump beams) are frequency detuned by $\delta v$. In this FWM process annihilates a pump photon from the field $E_1$ and the target photon at the signal frequency $v_s$ and creates photons on field $E_2$ and at the idler frequency $v_i = v_s + \delta v$ such that the total energy is conserved. A critical aspect of BS-FWM as compared to conventional FWM is that it provides a coherent/phase sensitive coupling between the two frequencies without adding noise and thus creates a bichromatic photon qubit. In the ideal case of a perfectly phase-matched process, the coherent coupling induced by the FWM-BS is expressed by the following coupled equations for the annihilation operators $\hat{a}_s(z)$ and $\hat{a}_i(z)$ for the signal and idler fields, respectively:

$$
\begin{pmatrix} \hat{a}_s(z=L) \\ \hat{a}_i(z=L) \end{pmatrix} = \begin{pmatrix} \cos(2\gamma PL) & i\exp(i\theta)\sin(2\gamma PL) \\ i\exp(-i\theta)\sin(2\gamma PL) & \cos(2\gamma PL) \end{pmatrix} \begin{pmatrix} \hat{a}_s(z=0) \\ \hat{a}_i(z=0) \end{pmatrix} \tag{1}
$$

where $\gamma$ is the nonlinear coefficient, $L$ is the propagation distance, and $P$ is the $|E_1|^2 = |E_2|^2$ is the power of the two pump beams with a relative phase $\theta$. From Eq. 1, it is seen that BS-FWM produces a rotation in the frequency Hilbert space $\{|v_s\rangle, |v_i\rangle\}$ such that $\gamma PL = \pi/8$ corresponds to a $\pi/2$ rotation. In addition Eq. 1 exhibits a phase dependence on the relative phases between the pump fields so that care must be taken to keep phase relationship between subsequent BS-FWM processes. Departure from the ideal case of noiseless unity-efficiency process, such as effects of imperfect phase matching, higher-order BS and spurious sources of noise, are discussed in the supplementary material.

A controllable relative phase between two optical frequencies (C) can be readily produced by propagating the bichromatic photon qubit over a length-tunable delay line. Measurement of the final frequency of the bichromatic photon (D) simply requires separating the two spectral components into two paths using a dispersive element and detecting in which path the photon is present using single photon detectors.

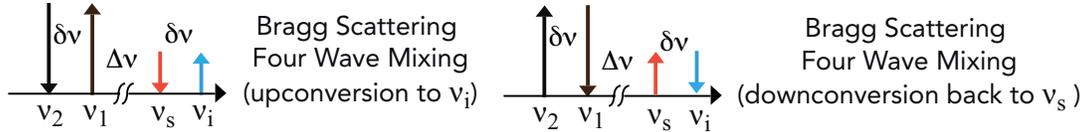

Figure 2. Up- (right) and down- (left) conversion via Bragg scattering four-wave mixing using one particular set of frequencies $\{v_1, v_2, v_i, v_s\}$. $\Delta v$ is the difference $|v_1 - v_s|$ between the nearest pump and single photon frequencies and $\delta v$ is the difference $|v_1 - v_2| = |v_i - v_s|$ corresponding to the shift in frequency.

**Toolbox for frequency generation and manipulation of a bichromatic qubit**

The full experimental setup is shown in Fig. 3 and is divided into 4 primary components A-D linked together to form the photonic Ramsey interferometer. Inset (A) shows the state preparation in which a single photon at $\lambda_s = 1283.8$ nm is heralded by detection with a silicon avalanche photodiode (Si-APD) of its partner photon generated at $\lambda_{\text{heralding}} = 940$ nm by type-0 spontaneous downconversion from a PPLN crystal pumped by a CW laser emitting at 543 nm. The heralding photon is spectrally filtered to select a signal photon at 1283 nm to a bandwidth of 0.5 nm. The heralded photon is separated from the residual pump and then coupled into an optical fiber to be routed to the second part of the setup. To align, synchronize, and characterize the setup, we use a tunable laser with a variable attenuator to produce a weak coherent state field with less than 0.1 photon per gate. The signal wavelength and bandwidth are selected to accommodate the frequency converter (inset B), and the $2^{\text{nd}}$-order correlation $g^{(2)}(t=0)$ of our single photon source is 0.23.  The design of the frequency converter is discussed in depth in the supplementary section (*S1-S6*) and is depicted in inset (B) of Fig. 3. In our setup, we employ a fiber-based BS-FWM scheme to couple two frequency within the same optical band (O-band) using only standard telecommunication components. The nonlinear medium is a dispersion-shifted fiber (DSF) with a measured zero-group velocity dispersion wavelength $\lambda_0 = 1421$ nm (fig. *S3*).  To minimize Raman noise, the fiber is cooled down to cryogenic temperature (fig. S1). Pump pulses are generated on demand by current modulation of laser diodes emitting at $\lambda_1 = 1551.7$ nm and $\lambda_2 = 1558.1$ nm. The resulting nanosecond pulses are amplified to multi-Watt level via cascaded erbium-doped fiber amplifiers, synchronized and overlapped with the signal photon in the DSF using wavelength division multiplexer add and drop filters. The polarization of all three fields is aligned to be parallel. For a signal at $\lambda_s = 1283.5$ nm, we first verified the power dependence of the conversion efficiency as shown in fig. 4 using a decoy state source with an average of 0.1 photon/pulse. The conversion scales as a square sine that mimics a Rabi oscillation. After correction for wavelength-dependent loss, the conversion efficiency reaches 90 ± 5% limited by power fluctuations of the pump beam and by higher-order leakage into other frequency modes due to the competing up-conversion process and is most apparent at higher pump powers where there is a clear discrepancy between the expected coherent oscillation and the observed result.

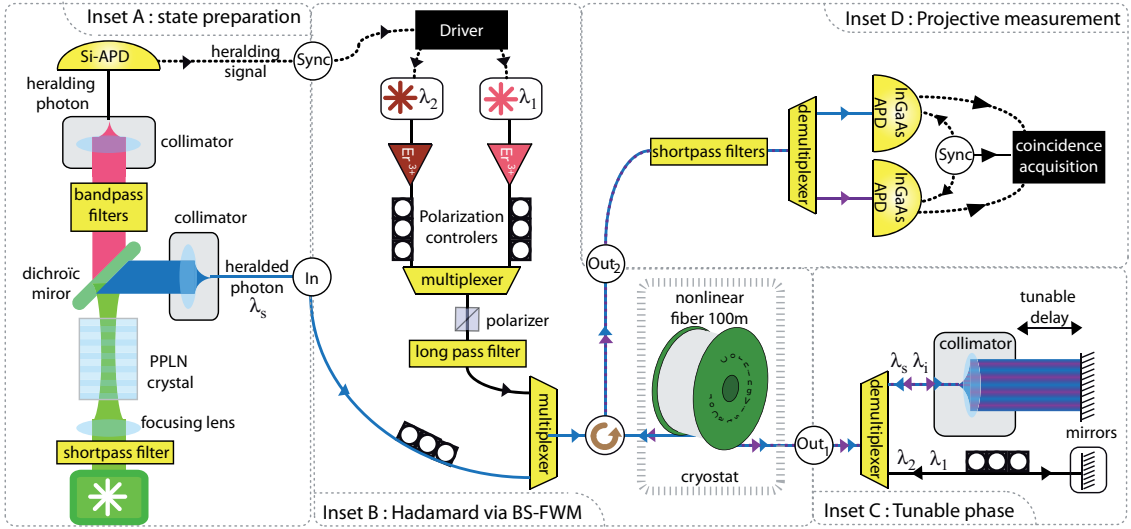

Figure 3. Insets A-D: Experimental implementation of Ramsey interferometry with photons.

To produce a tunable phase between the two states, we filtered the pump beams from the bichromatic photon and inserted a tunable free space delay on its path [inset (C) of fig. 3]. As indicated by Eq. 1, the BS-FWM process depends not only on the relative phase between the two spectral components of the single photon but also on the relative phase between the two pump fields. It is important to preserve the phase relationship between pumps and signal for driving a subsequent BS-FWM interaction, and thus we must control independently the relative phase between pump and the bichromatic qubit since otherwise the precession acquired by the two pumps ($\theta$ in eq. S3) would exactly cancel out with the bichromatic qubit phase $\varphi = 2\pi\Delta x(v_s - v_i)/c$ which accumulated in propagating over a distance $\Delta x$.

The inset (D) of fig. 3 illustrates how the projective measurements are performed. First, the remaining photons from the optical pump in frequency converters are filtered out using short pass filters, and the two spectral components of the bichromatic qubit are separated using commercially available wavelength division demultiplexers. The projective measurement is then made by performing single-photon detection on each arm using InGaAs avalanche photodiodes. Since our single-photon detectors can operate only in a gated mode, they are synchronized to match the arrival time of the single photons (typically using the heralding event). The photon flux and correlations at the detectors are determined using a time-tagging module (coincidence acquisition).

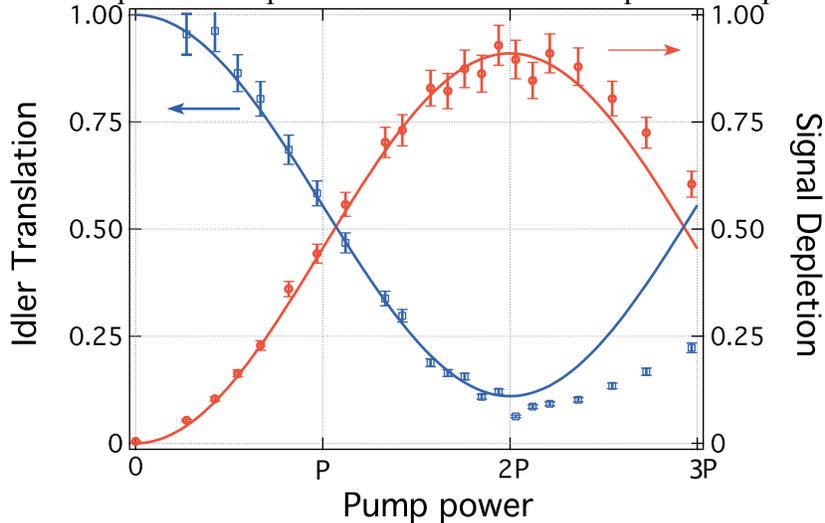

Figure 4. Efficiency of the BS-FWM as a function of the pump power where P is the power corresponding to a π/2 rotation on the Bloch sphere. This plot is the Rabi oscillation between the two energy levels of the bichromatic qubit and is obtained using the setup depicted in fig. S6 (Inset B).

**Ramsey interference**

Our approach allows generating any state of a bichromatic qubit and thus setting the bichromatic state of the photon anywhere on the Bloch sphere. The setup is depicted in fig. 3 and consists of the four components discussed above. The photon is prepared in frequency state $|v_s>$ (A), a π/2 pulse is applied using the quantum frequency converter (B), a tunable phase is imparted on the bichromatic qubit (C), a π/2 pulse is again applied using the same frequency converter (B) but in the reverse direction, and lastly the final state is reconstructed via frequency demultiplexing and single-photon detection (D). In our experiments, the BS-FWM pump power $P$ is adjusted to give a conversion efficiency of 1/2 ($P$ = 2 watts) so that the bichromatic qubit exits (B) as a balanced superposition of the two frequencies $v_s$ and $v_i$. In the tunable phase delay stage (C), the pump and signal arms are kept of nearly equal length so that the optical pulses temporally overlap with the reflected single photon when they are combined back in the DSF. The resulting Ramsey interference is depicted in fig. 5 which shows the signal conversion to the idler frequency as a function of the imparted phase. As expected, the interference pattern exhibits fringes corresponding to the probabilities $p(v_s) = \sin^2(\varphi/2)$ and $p(v_i) = \cos^2(\varphi/2)$. The interference pattern shows fringes over a $\pi$ phase that corresponds to free-space propagation of 0.36 mm. The interference fringes are the proof that the underlying BS-FWM process preserves the coherence of the quantum fields. The visibility of the fringes is nearly 50% and is limited by the following two factors. The bandwidth of the single photon is nearly equal to the acceptance bandwidth of the BS-FWM, which limits the maximum conversion to 80% (see fig. S3). We verified this by using a coherent-state signal field with an input bandwidth better matched to that of the BS-FWM and we observe improved visibility > 80% [Put in the actual values.] (see

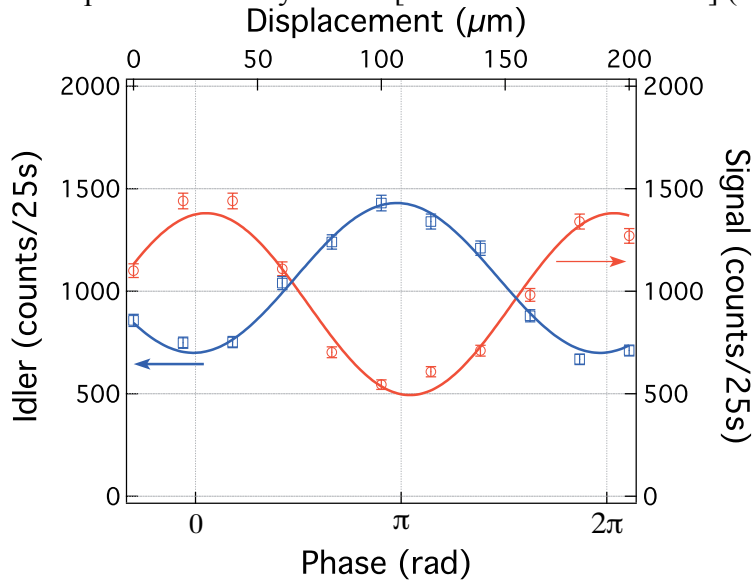

Figure 5. Ramsey interference fringes observed while varying the phase between two BS-FWM frequency converters.

fig. S3). The second factor that limits the visibility is the loss of 2.5 dB experienced by the pump field in the tunable delay element, which results in the second rotation being less than $\pi/2$ as targeted.

We have also verified that our system is suitable for quantum information applications and preserves Fock states (i.e., technical noise does not spoil the fidelity), by measuring the second-order correlation function $g^{(2)}(t = 0)$ of the output photon. This is achieved by replacing the demultiplexer by a balanced beamsplitter in the projective measurement (D). We find that independent of whether we apply a transformation on the Bloch sphere or not (pump beams turned off), the $g^{(2)}(t = 0)$ function remains at a value of 0.2.

**Conclusion**

We have introduced a complete set of building blocks for encoding, manipulating and measuring quantum information contained in frequency qubits. We have demonstrated a single photon can be placed in a bichromatic state at any point on the corresponding Bloch sphere using a photonic Ramsey interferometer. As a proper interferometer, it is important to note this setup does not require stability down to the wavelength range but only at the larger beating wavelength $\Delta\lambda = c/\delta\nu$.

We believe our demonstration will find numerous applications in quantum information. Indeed bichromatic photons can constitute the interface between quantum systems operating at different frequencies (26-28) such as quantum information carrier and quantum repeaters. A bichromatic photon can also serve as a stable quantum information carrier. A potential extension of our work is the manipulation of entangled states rather than pure states. Photon pairs spontaneously generated on frequency combs (*31*) constitute an example of high dimension frequency entangled state that could be manipulated using our Ramsey interferometer to perform quantum key distribution with improved robustness (*32,33*). In addition, we foresee that single photon spectral-temporal pulse shaping (*34*) using nearly identical setup as our Ramsey interferometer is another promising application as it may also serve as an interface between bandwidth-time unmatched quantum optics systems. Both these aspects are highly relevant to quantum key distribution (*35-38*) whose extension to longer distances will depend on quantum relays and overall robustness again loss and noise.

Beyond these purely quantum information outcomes, we think they are important applications that will make use of the interferometer itself. Indeed, our demonstration enables measuring spectrally dependent phase change, i.e. perform quantitative phase spectroscopy (*39-40*), with a very low amount of light. That is particularly relevant for performing spectroscopy on samples that are photosensitive and/or cannot tolerate any absorption. Biological samples such as eyes or phototrophs (organisms carrying out photosynthesis) may be studied using our single photon Ramsey interferometer. For the later case, the quantum coherence of the probe light may even help studying the quantum nature of the photosynthesis process (*41-42*)

source of heralded single photons", *Opt. Expr.* **22**, 6535 (2014)

**Acknowledgments:**

We acknowledge support from the Defense Advanced Research Projects Agency via the QuASAR program and the Air-Force Office of Scientific Research under grant FA9550-12-1-0377. Sven Ramelow is funded by a EU Marie Curie Fellowship (PIOF-GA-2012-329851).

**Supplementary Materials:**

**S1. High efficiency and low noise BS-FWM**

For the Ramsey interference to have a good visibility it is important that the frequency transfer can be complete, i.e. it must be possible to impart a $\pi$ polar rotation on the Bloch sphere. The BS-FWM efficiency is given by

$$\eta = \left| \frac{i\kappa}{\sqrt{|\kappa|^2 + |\delta k|^2}} \sin\left(\sqrt{|\kappa|^2 + |\delta k|^2}\, L\right) \right|^2 \qquad (S1)$$

where $\kappa = 2\gamma E_1 E_2^* = 2\gamma Pe^{i\theta}$ quantifies the nonlinear interaction over a distance L for $|E_1|^2 = |E_2|^2 \equiv P$ and $\delta k$ is the phase mismatch. From this expression, it is clear that coherent oscillation with maximum amplitude arises when the phase mismatch vanishes and with a period given by the product $\kappa L$. A full frequency conversion arises when $\kappa L = \pi(1/2+n) \ \forall n \in \mathbb{N}$ so that a $\pi$-rotation on the Bloch sphere requires $\kappa L = \pi/2$.

If the phase matching is not satisfied, i.e. the phase mismatch $|\delta k|$ is comparable or greater than $|\kappa|$ then the coherent oscillation is of larger frequency and lower amplitude. It is interesting to note, that the phase mismatch $\delta k$ in the Bragg scattering process plays the same role as the frequency detuning to the Larmor and Rabi frequencies for RMN and Ramsey interferometry respectively.

Assuming phase matching is indeed satisfied (see section S5), the evolution of the signal and idler fields due to the Bragg scattering process is given by the coupled equations:

$$\frac{\partial}{\partial z}\begin{pmatrix} \hat{a}_s(z) \\ \hat{a}_i(z) \end{pmatrix} = \begin{pmatrix} 0 & i\kappa \\ i\kappa^* & 0 \end{pmatrix}\begin{pmatrix} \hat{a}_s(z) \\ \hat{a}_i(z) \end{pmatrix} \qquad (S2)$$

whose analytic solution is a rotation in the frequency Hibert space $\{|v_i\rangle, |v_s\rangle\}$

$$\begin{pmatrix} \hat{a}_s(z=L) \\ \hat{a}_i(z=L) \end{pmatrix} = \begin{pmatrix} \cos(2\gamma PL) & i\exp(i\theta)\sin(2\gamma PL) \\ i\exp(-i\theta)\sin(2\gamma PL) & \cos(2\gamma PL) \end{pmatrix}\begin{pmatrix} \hat{a}_s(z=0) \\ \hat{a}_i(z=0) \end{pmatrix} \qquad (S3)$$

From eq. S3, the evolution of $\hat{a}_s(z)$ is phase sensitive when $\hat{a}_s(0) \neq 0 \neq \hat{a}_i(0)$, i.e when the state isn't on the poles of the Bloch sphere. It should be noted that not only the phase $\varphi$ between $\hat{a}_s(0)$ and $\hat{a}_i(0)$ plays a role but also the phase $\theta$ between the two pump fields.

The process of Bragg scattering four wave mixing (BS-FWM) is in principle noiseless. Unlike other four-wave mixing processes, it annihilates the signal photon to create an exact copy that only differs through its frequency. In addition, the process can theoretically reach 100% efficiency. However, eq. S1 is a good description of the frequency conversion via BS-FWM only if it does not compete with any other processes that could either annihilate the photon at the frequency $v_s$ or $v_i$ or create a noisy photon at those same frequencies. The first effect would reduce the possible angle of the polar rotations on the Bloch sphere while the second would break down the assumption of an ideal two level state.

Practically, high efficiency and low noise are not trivial to achieve together. The main reason for that is that other nonlinear processes may arise and compete with the BS-FWM. In our implementation, we have identified two sources of noise: spontaneous Raman scattering and spontaneous four wave mixing. We also have identified two sources of reduced efficiency: Raman anti-Stokes absorption and a competing BS-FWM process.

## S2. Spontaneous four-wave mixing

Spontaneous four wave mixing is a process that annihilates two photons from the pump at frequencies $v_j$, $v_k$ (with $j, k \in [1, 2]$) to create two new photons at frequencies $v_s$ and $v_i$ such that $v_k + v_j = v_i + v_s$. If either $v_i$ or $v_s$ correspond to the frequency of the signal or idler photon in our BS-FWM process, then photon noise will be added to the process. Fortunately, the phase matching for this process is far different from the one for BS-FWM. At 0 Kelvin, spontaneous four-wave mixing is the only source of noise we expect. Its spectrum is depicted in fig. S1 (blue curve only) for a single pump at 1550 nm. Out of a narrow phase matching band close to the pump wavelength (not illustrated), the photon flux decays with the frequency detuning as the inverse of $\beta_2(2\pi\Delta\nu)^2$. To avoid spurious photons at the wavelength we desire manipulating our bichromatic qubit, we have to make sure the group velocity dispersion ($\beta_2$) at the pump frequencies $\nu_1$ and $\nu_2$ is significant and that the frequency detuning $\Delta\nu$ between the pump beams and the bichromatic qubit is large enough.

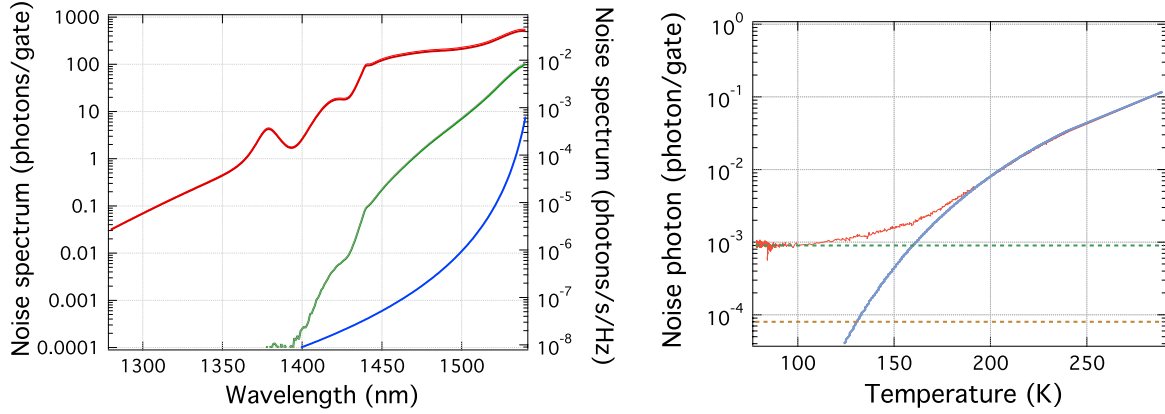

Figure S1: Left: Theoretical prediction of the spontaneous anti-Stokes noise spectrum originating from spontaneous four wave mixing and Raman scattering at 300 K (red) 100 K (green) and 0 K (blue). Right: Experimental noise characterization versus temperature.

## S3. Raman scattering

Raman scattering in optical fibres has been studied extensively and it is well known that spontaneous Raman scattering is a source of noise for photon pair generation in silica fibres. In silica fibres, Raman scattering can be greatly suppressed on the anti-Stokes (blue shifted) side of the spectrum by cooling the fibre or by having frequency detuning much larger than the peak of the Raman response at 15 THz. Figure S1 shows the theoretical Raman spectrum at room temperature and at liquid nitrogen temperature. It shows clearly that we need to maximize the detuning $\Delta\nu$ and cool the fibre to reduce the phonon population and thus the to reduce the photon noise. Our predictions are plotted as a fraction of photon per detection gate which corresponds to an integration time of 4 ns (the pump pulse duration) and a bandwidth integration of 12 nm (the bandwidth of our spectral filter) and in term of spectral brightness (photon/Hz/s).

We have measured the photon noise at the bichromatic photon wavelength of 1300nm as a function of the temperature of our nonlinear fibre. The minimum noise we achieved was $10^{-3}$ which was limited by small portion of fibre (around 2 meters) that we couldn't cool down because of technical limitations.

## S4. Competing BS-FWM processes

While the BS-FWM can reach 100% efficiency, this is only true when the system can indeed be described by only the four frequencies $\{\nu_1, \nu_2, \nu_s, \nu_i\}$ involved in the up-conversion BS-FWM process such that $\nu_i = \nu_s + |\nu_1 - \nu_2|$ . The same set of initial frequencies $\{\nu_1, \nu_2, \nu_s\}$ can actually be involved in a down-conversion process thus creating a different idler frequency $\nu_d = \nu_s - |\nu_1 - \nu_2|$ (see fig. S2). If both processes are phase matched, then the photon will experience the up-conversion or down-conversion process with equal probability thus effectively reducing the up-conversion probability of converting a single photon from its initial frequency $\nu_s$ to its final frequency $\nu_i$ to 50%.

In addition to that, higher order BS-FWM may also occurs where the photon up-converted to $v_i$ is subsequently up-converted rather than down-converted back to its initial frequency $v_s$.

In both cases, the solution is to make sure than only the BS-FWM process of interest is phase matched. That is achieved by a tight phase matching condition that we are detailing in the coming paragraph.

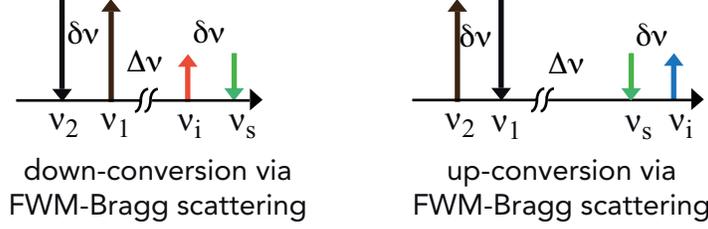

Figure S2. Bragg scattering up and down conversion processes.

## S5. Phase matching

From the above analysis, it is clear it is preferable to have the two pump frequencies at longer wavelength than the bichromatic photon. The BS-FWM configuration for the pump and signal frequencies is illustrated by fig. S2 . In this configuration, Raman scattering and spontaneous four-wave mixing are suppressed by having $\Delta v > 20$ THz, while the down-conversion BS process is suppressed by having $\delta v$ large enough. The phase matching for the up-conversion process illustrated in fig. S2 is given by

$$\delta k = \beta_{2,v_0}\left(\delta\omega^2 + \delta\omega\Delta\omega\right) + \frac{\beta_{4,v_0}}{12}\left(\delta\omega^4 + \delta\omega\Delta\omega\left(\frac{3}{2}\delta\omega\Delta\omega + \frac{\Delta\omega^2}{2} + 2\delta\omega^2\right)\right) \qquad (S4)$$

where $\Delta\omega$ is the pump-photon spectral detuning (in angular frequency) and $\delta\omega$ is the detuning between the two pump in angular frequencies. In first approximation, phase matching is ensured by having zero group velocity dispersion (GVD) at the average wavelength $v_0 = \frac{1}{4}(v_s + v_i + v_1 + v_2)$ (note: that the phase matching condition for the up-conversion process can be expressed the same way but for the central frequency $v_0 = \frac{1}{4}(v_s + v_d + v_1 + v_2)$). As the forth order term of the dispersion doesn't vanish, $\delta\omega$ and $\Delta\omega$ has to be reasonably low to keep the phase matching. There is thus a trade-off between good phase matching of the desired BS-FWM up-conversion process and high phase mismatching of the undesired BS-FWM down-conversion process. Furthermore, a too tight phase matching condition limits the acceptance bandwidth of the BS-FWM process, which is practically problematic for typical single photon sources that are not particularly narrowband.

## S6. Design of the experiment

Accounting for all the constraints we have made explicit above, we have chosen to have our pump beams in the telecom C-band (1530 to 1565 nm) where Erbium doped amplifier are easily available while the bichromatic qubit lies in the telecom O-band (1260 to 1360 nm ; also renamed as *Quantum-band* by various scientists working on quantum key distribution). This configuration gives a frequency detuning of 40 THz (260 nm in term of wavelength detuning) much more than the 13-15 THz of the Raman spectral peak in silica.

We have selected the Corning VistaCorTM for its zero-group velocity dispersion (GVD) point located half way between the C and O-bands. The dispersion has been estimated by a time of flight measurement depicted in fig. S3. We used 3 tunable laser covering a broad wavelength range (1260 to 1640 nm). The CW laser beam is carved into nanosecond pulses by an electro-optic modulator and sent to the fibre to be characterized. The time of flight is measured with an oscilloscope that acquires the pulse train detected by a photodiode and is triggered by the driver of the electro-optic modulator. The result of our measurement is displayed in fig. S3. The curves shows a zero-GVD at 1421 nm at room temperature. At cryogenic temperature, the zero GVD is slightly shifted to shorter wavelength which we deduced from the modified phase matching condition for the BS-FWM process. From this fibre and pump wavelength in the C-band, our bichromatic qubit can be manipulated in the frequency interval 1280-1310 nm. This has been verified by having tunable lasers as the pump beams and another tunable laser as the signal photon source. With that target wavelength, we have designed a heralded photon source that delivers single photons at 1295 nm. The source based on spontaneous down conversion in a PPLN crystal is described in the main text of our work. The emitted flux of photon has to be spectrally filtered to make its bandwidth compatible with the acceptance bandwidth of BS-FWM. The filtering is done on the heralding arm with the detrimental effect of a larger photon noise on the heralded signal. The BS-FWM efficiency has been measured with the heralded single photon source and it is plot in figure S3. The conversion efficiency is reduced to 75% as compared to 90% with photons originating from narrowband faint pulses. This is mainly due to the bandwidth missmatch between the single photon and the phase matching condition.

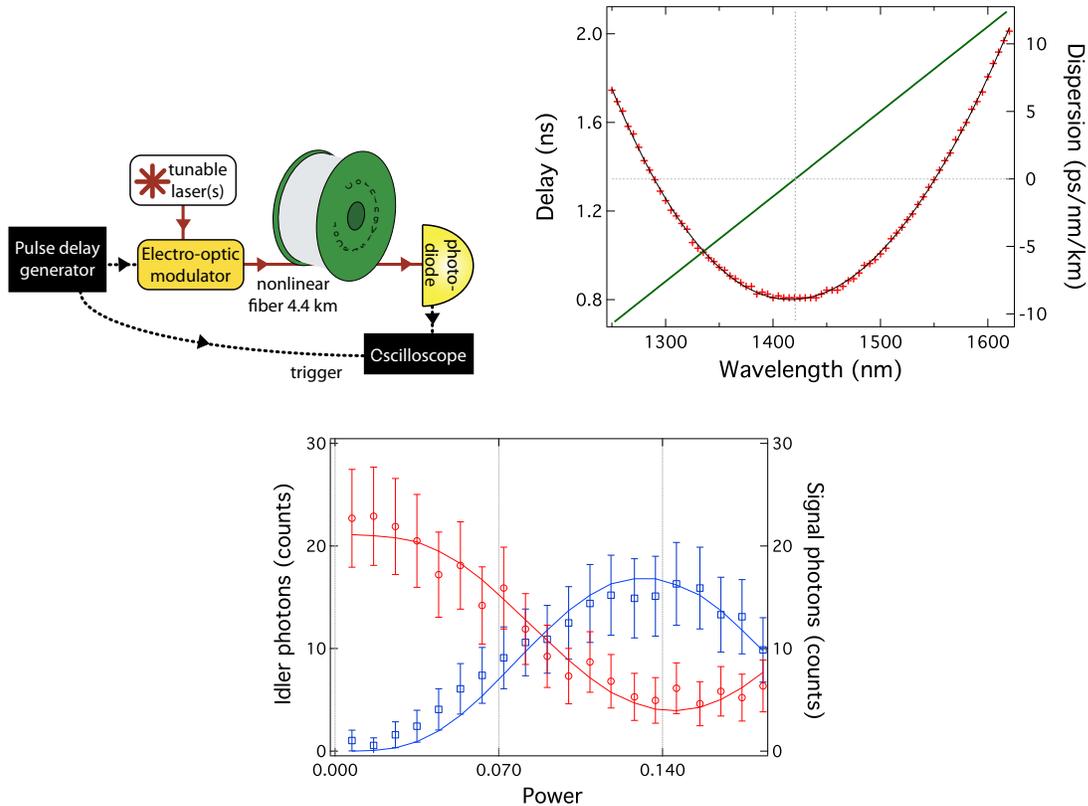

Figure S3. Left: Dispersion measurement via time of flight measurement. Center: Dispersion curve for Corning VistaCorTM fibre. Right: BS-FWM efficiency as a function of power using single photons originating from the PPLN heralded source.

**S7. Ramsey measurement for measuring variation of group velocity dispersion with temperature**

In this experiment we want to estimate the variation of the group velocity dispersion of a short piece of fibre as a function of temperature changes. The experimental setup is similar to the one introduced in fig. S4 but differs in a few respects. Here the photon source is a nanosecond pulse attenuated down to 0.1 photon/pulse. The signal, idler and pump wavelength are different $\lambda_s =$ 1293.8 nm, $\lambda_i = 1305$ nm, $\lambda_1 = 1536$ nm, $\lambda_2 = 1550.6$ nm. In this setup, the two $\pi/2$ pulses are imparted by BS-FWM in two different spool of fibre that are not cooled down to cryogenic temperature. The second nonlinear fibre is slightly longer to compensate for the excess loss experienced by the pump between the first and the second $\pi/2$ pulses. The frequency detection is here made with a single avalanche photodiode, a dispersion compensation module (DCM) that temporally separates the spectral components of the bichromatic photon and a time tagging module that measures its time of arrivals. Those 3 elements act thus as an optical spectrum analyser operating down to the single photon level. The biggest difference of this setup against the one present in figure 3 is that the phase is acquired by both 4 fields. The pump beams co-propagates with the bichromatic photon in the fibre to be measured. As we said earlier, if both 4 fields propagate together in free space, the imparted phase vanishes. However, this is not true if the medium the fields propagate through posses group velocity dispersion. The relative phase accumulated in that case is:

$$\phi = \phi\left(\nu_s\right) + \phi\left(\nu_1\right) - \phi\left(\nu_i\right) - \phi\left(\nu_2\right) \quad \text{where} \quad \phi\left(\nu_j\right) = 2\pi\nu_j t - \beta\left(\nu_j\right)L$$

For varying temperature, the individual and total phases vary as

$$\frac{\partial \phi\left(\nu_j\right)}{\partial T} = \frac{\partial \beta\left(\nu_j\right)}{\partial T}L + \beta\left(\nu_j\right)\frac{\partial L}{\partial T} \Rightarrow \frac{\partial \phi\left(\nu_j\right)}{\partial T} = 4\pi^2\left(\delta\nu\left(2\Delta\nu + \delta\nu\right)\right)\left(\frac{\partial \beta_2\left(\nu_0\right)}{\partial T}L + \beta_2\left(\nu_0\right)\frac{\partial L}{\partial T}\right)$$

where we have Taylor expanded the wavenumbers in $\Sigma_j \beta(\nu_j)$ around $\nu_0 = (\nu_s + \nu_i + \nu_1 + \nu_2)/4$ and we call $\Delta\nu$ the detuning between $\nu_0$ and $\nu_s$ and $\delta\nu$ the detuning between $\nu_s$ and $\nu_i$. In the present case the group velocity dispersion is in the range of $\beta_2(\nu_0) \approx 8 \times 10^{-27}$ s$^2$/m (SMF-28 from Corning) and the thermal coefficient of silica $\partial L / \partial T = 5.5 \times 10^{-7}$ m/K.

The result of our experiment is plotted as an inset of fig. S4 showing Ramsey fringes with a period of 120 K. In the expression of $\partial \phi / \partial T$, the second term of the sum is negligible in comparison to the first so that we have

$$2\pi = \frac{\partial \phi}{\partial T}\Delta T = 4\pi^2\left(\delta\nu\left(2\Delta\nu + \delta\nu\right)\right)\frac{\partial \beta_2\left(\nu_0\right)}{\partial T}L\Delta T$$

As $\Delta T = 120$K, $L \approx 30$ m, $\Delta\nu = 20$ THz and $\delta\nu = 0.8$ THz, we conclude that $d\beta_2/dT = 0.0014$ps$^2$/km/K which corresponds to a shift of the zero-dispersion wavelength by almost 2 nm for a heating by 120 K.

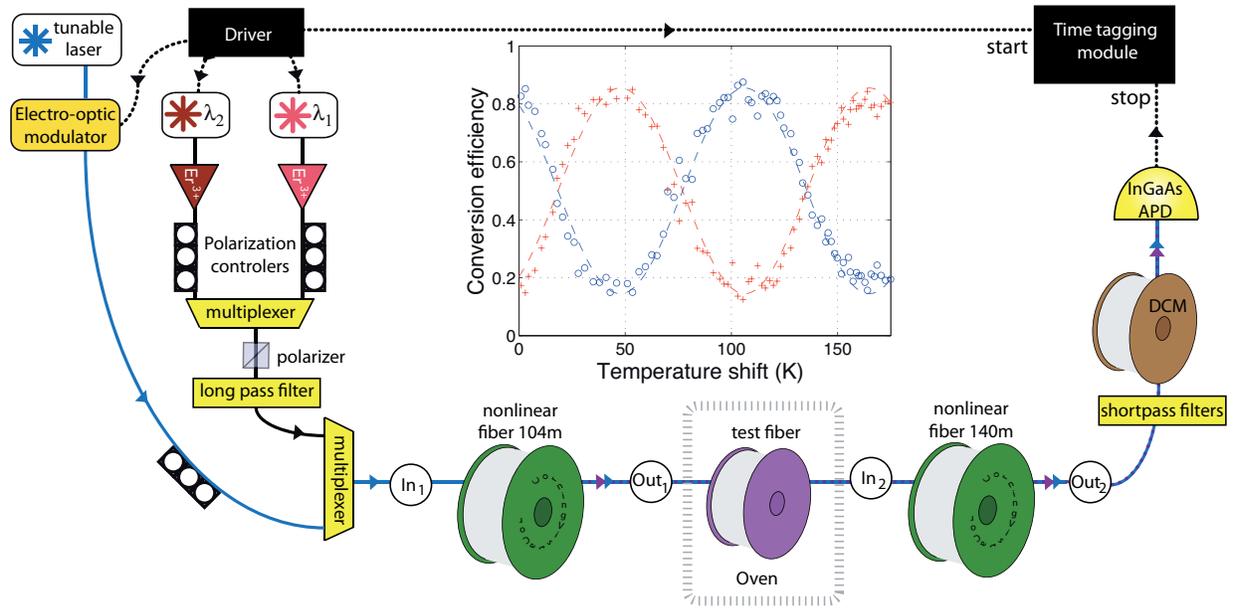

Figure S4. Measurement of the group velocity dispersion with a Ramsey interferometer. Inset: fringes resulting from the experiment.